\documentclass[%
  reprint,
  superscriptaddress,
  nofootinbib,
  amsmath,amssymb,
  aps,
  prd
]{revtex4-1}

\usepackage{graphicx}
\usepackage{dcolumn}
\usepackage{bm}

\usepackage{graphicx,amsmath,amssymb,soul,bbm,dcolumn, multirow,array,float,enumitem,lmodern,dsfont,microtype,nicefrac}
\usepackage[dvipsnames]{xcolor}
\usepackage[utf8]{inputenc}
\def\fm {\mathop{\hbox{fm}}}
\def\MeV {\mathop{\hbox{MeV}}}

\def\Re {\mathop{\hbox{Re}}}

\def\Tr {\mathop{\hbox{Tr}}}

\def\beq{\begin{equation}}
\def\eeq{\end{equation}}
\def\beqs#1\eeqs{\beq\begin{split} #1 \end{split}\eeq}

\def\dd#1#2{\frac{d #1}{d #2}}

\def\comment#1{}

\def\av#1{ \left\langle #1 \right\rangle }

\def\chipt{$\chi$PT\ }

\usepackage{array}   
\newcolumntype{L}{>{$}l<{$}} 
\newcolumntype{S}{>{\footnotesize $}l<{$\normalsize}} 
\usepackage{ctable} 

\usepackage[colorlinks=true,backref=section,linktocpage=true,
  citecolor=NavyBlue,urlcolor=NavyBlue,linkcolor=NavyBlue,pdfpagemode=UseOutlines]{hyperref}
\hypersetup{%
  bookmarksnumbered=true,
  pdftitle = {},
  pdfsubject = {},
  pdfauthor = {},
  pdfkeywords = {}
}

\usepackage{hypernat} 

\begin{document}

\title{Setting the scale for nHYP fermions with the L\"uscher-Weisz gauge action}

\author{Hossein Niyazi}
\email{hosseinniyazi@gwu.edu}
\affiliation{Department of Physics, The George Washington University, Washington, DC 20052, USA}
\author{Andrei Alexandru}%
\email{aalexan@gwu.edu}
\affiliation{Department of Physics, The George Washington University, Washington, DC 20052, USA}
\affiliation{Department of Physics, University of Maryland, College Park, MD 20742, USA}

\author{Frank X. Lee}
\email{fxlee@gwu.edu}
\affiliation{Department of Physics, The George Washington University, Washington, DC 20052, USA}
\author{Ruair\'i Brett}
\email{rbrett@gwu.edu}
\affiliation{Department of Physics, The George Washington University, Washington, DC 20052, USA}

\date{\today}

\begin{abstract}
  Lattice QCD calculations using gauge smearing for fermion kernels are computationally efficient. Hypercubic blocking (nHYP smearing) has been shown to reduce scaling errors. In this work we use an improved action for $N_f=2$ QCD, based on the L\"uscher-Weisz gauge action and clover-improved Wilson fermions with nHYP smeared gauge links. We perform a parameter scan in the region with lattice spacing between $0.066 \fm$ and $0.115 \fm$ and pion mass between $207 \MeV$ and $834 \MeV$.
  We determine the lattice spacing and pion mass as a function of the bare coupling parameters~($\beta$ and~$\kappa$). The results are obtained from twenty-two ensembles on a $24^3\times 48$ lattice to percent level in statistical accuracy. The finite-volume effects for these ensemble are at the sub-percent level. From these measurements we produce easy-to-use parameterizations to help tune simulations with this action. The lattice spacing is fixed using a mass-independent procedure, by matching observables in the chiral limit. We also provide a parameterization for the chiral extrapolation which is universal and should hold for all discretizations of $N_f=2$ QCD. 
\end{abstract}

\maketitle


\section{\label{sec:intro}Introduction}

Quantum Chromodynamics (QCD) is the fundamental theory of the strong interaction, with only a few parameters: the strong coupling constant ($g$) for gluon-gluon and gluon-quark interactions, and the quark masses ($m_q)$ (one for each quark flavor). In the energy region relevant for hadronic interactions the effective coupling is strong. Lattice QCD can be used to calculate QCD properties in this region via numerical simulations on a spacetime lattice with spacing~$a$. 
Lattice QCD allows us to extract physical quantities in lattice units (or in dimensionless ratios) from Euclidean correlation functions.
To relate such dimensionless quantities to physical quantities in physical units we must determine the lattice spacing.
In practice, this is done by fixing one physical observable to its experimental value. 
Once the lattice spacing is fixed, the values of other observables can be converted to physical predictions.
For this reason, scale setting is an important step in lattice QCD calculations. 

Lattice QCD calculation are always done at finite lattice spacing. All predictions need to be extrapolated to the continuum limit, to eliminate the discretization errors. Generically discretizations errors are expected to vanish linearly with the lattice spacing as we take the continuum limit, but improved actions can be used to accelerate the convergence rate. The largest discretization errors are usually generated by the quark contribution to the action. Smearing the gauge fields was found to improve the scaling of hadron spectrum significantly~\cite{Albanese:1987ds,Hasenfratz:2001hp,Alexandru:2004ge,D_rr_2009,D_rr_2011}. With the introduction of analytical smearing~\cite{Morningstar_2004}, efficient dynamical simulations were made possible using Hybrid Monte Carlo~(HMC) method~\cite{Duane:1987de}. Hypercubic blocking~\cite{Hasenfratz:2001hp} was proposed to reduce scaling violations and taste breaking effects for staggered fermions and an analytic version was designed, nHYP, that works with HMC~\cite{Hasenfratz:2007rf,Hoffmann:2007nm}. This action is used by a number of lattice QCD groups, including ours, for studies of hadronic physics~\cite{Hasenfratz:2007rf,Hoffmann:2007nm,Lang:2011mn,Prelovsek:2011im,Pelissier:2012pi,Prelovsek:2013ela,Lang:2014tia,Freeman_2014,Lujan_2016,Guo:2016zos,Guo:2018zss,Culver:2019qtx}.

In this work, we consider an action for QCD with two mass-degenerate quark flavors using nHYP clover fermion discretization~\cite{Hasenfratz:2007rf} for quark fields and L\"{u}scher-Weisz gauge action~\cite{Luscher:1984xn}. 
We study the scale setting for this action to facilitate its wider usage.
The plan of the paper is the following:
In Sec.~\ref{sec:action}, we give a largely self-contained description of the action. In Sec.~\ref{sec:lsp} we present a brief review of the methodology for scale setting, our numerical simulation results, and finite volume effects. 
Smooth parametrization functions are extracted in Sec.~\ref{sec:param}, before conclusions in Sec.~\ref{sec:sum}.

\section{\label{sec:action}Action}


The lattice QCD action we use in this study is a function 
of the gauge links $U_\mu$ and quark fields $\psi$, 
\beqs
S(U,\psi)&= S_\text{gauge}(U) + S_\text{quark}(U) \\&= S_\text{LW}(U) + \sum_f \bar\psi_f D_\text{clover}(\kappa, V) \psi_f \,.
\eeqs
Note that the gauge part of the action $S_\text{LW}(U)$ is
a function of the {\em thin} links, whereas the links $V$
that enter the fermionic matrix $D_\text{clover}$
are nHYP smeared links $V(U)$. In this section we review the
details for each of the term in this action.

The L\"uscher-Weisz gauge action~\cite{Luscher:1984xn} $S_\text{LW}$
is an improved action designed to cancel $O(a^2)$ errors in the standard Wilson gauge action by adding two gauge-invariant operators beyond the usual plaquette,
\beqs
S_\text{LW} &=\beta \sum_x \Bigg[ \sum_{\mu<\nu} \frac{1}{3}\Re(1-\Tr U_{\mu\nu-\mu-\nu}(x)) \\
&+ c_\text{rt} \sum_{\mu\neq\nu} \frac{1}{3}\Re(1-\Tr U_{\mu\mu\nu-\mu-\mu-\nu}(x))  \\
&+ c_\text{pg} \sum_{\mu\nu\rho \in {\cal S}_\text{pg}} \frac{1}{3}\Re(1-\Tr U_{\mu\nu\rho-\mu-\nu-\rho}(x))\Bigg],
\label{eq:LW}
\eeqs
where we use the shorthand notation
\beq
U_{\mu_1\mu_2\cdots\mu_k}(x)\equiv 
U_{\mu_1}(x)U_{\mu_2}(x+\hat{\mu}_1)\cdots U_{\mu_k}(x+\sum_{j=1}^{k-1}\hat{\mu}_j).
\eeq
The summation indices run in both positive and negative directions $\{\pm 1,\pm 2,\pm 3, \pm 4\}$. For negative indices we have $U_\mu(x)\equiv U_{-\mu}(x+\hat{\mu})^\dagger$.
The set of indices in the parallelogram term has 16 elements,
\beqs
{\cal S}_\text{pg}=\big\{ &
(\mu,\nu,\rho), (\mu,-\nu,\rho), (\mu,\nu,-\rho), \\
&(\mu,-\nu,-\rho) \,\vert\, \mu<\nu<\rho
\big\}\,.
\eeqs
%
%
The three terms correspond to three types of gauge-invariant lattice loops:  square, rectangle, and parallelogram. These terms can be depicted graphically as,
\beqs
U_\text{pl} &= U_{\mu\nu-\mu-\nu} =
\setlength{\unitlength}{0.7pt}
\begin{picture}(25,25)(10,16)
  \put(10,30){\vector(1,0){12.5}}
  \put(10,10){\line(1,0){20}}
  \put(10,10){\vector(0,1){12.5}}
  \put(10,10){\line(0,1){20}}
  \put(30,10){\vector(-1,0){12.5}}
  \put(30,30){\line(-1,0){20}}
  \put(30,30){\vector(0,-1){12.5}}
  \put(30,30){\line(0,-1){20}}
\end{picture}\,,\\
U_\text{rt} &= U_{\mu\mu\nu-\mu-\mu-\nu} =
\setlength{\unitlength}{0.7pt}
\begin{picture}(45,25)(10,16)
  \put(10,10){\vector(0,1){12.5}}
  \put(10,10){\line(0,1){20}}
  \put(10,30){\vector(1,0){12.5}}
  \put(10,30){\vector(1,0){32.5}}
  \put(10,30){\line(1,0){40}}
  \put(50,30){\vector(0,-1){12.5}}
  \put(50,30){\line(0,-1){20}}
  \put(50,10){\vector(-1,0){12.5}}
  \put(50,10){\vector(-1,0){32.5}}
  \put(50,10){\line(-1,0){40}}
\end{picture}\,,\\
U_\text{pg} &= U_{\mu\nu\rho-\mu-\nu-\rho} =
\setlength{\unitlength}{0.7pt}
\begin{picture}(45,30)(10,20)
  \put(10,10){\vector(0,1){12.5}}
  \put(10,10){\line(0,1){20}}
  \put(10,30){\vector(2,1){10}}
  \put(10,30){\line(2,1){15}}
  \put(25.2,37.6){\vector(1,0){12.5}}
  \put(25.2,37.6){\line(1,0){20}}
  \put(45.2,37.6){\vector(0,-1){12.5}}
  \put(45.2,37.6){\line(0,-1){20}}
  \put(45.2,17.6){\vector(-2,-1){10}}
  \put(45.2,17.6){\line(-2,-1){15}}
  \put(30,10){\vector(-1,0){12.5}}
  \put(30,10){\line(-1,0){20}}
\end{picture} \,.
\eeqs
\medskip

The coefficients in the action are taken from one-loop, tadpole-improved perturbation theory~\cite{Alford:1995hw}
\begin{align}
\begin{split}
c_\text{rt} &= -\frac1{20 u_0^2}(1+0.4805\alpha_s) \,, \\
c_\text{pg} &= -\frac1{u_0^2}0.03325\alpha_s \,, \\
\alpha_s &= - \frac{4\log u_0}{3.06839} \,, \quad\quad 
u_0 = \left( \frac13 \Tr \langle U_\text{pl}\rangle \right)^{1/4} \,,
\end{split}
\label{eq:sg}
\end{align}
where the measured expectation value of the plaquette
is used to determine both the value of the mean link $u_0$ and the QCD coupling
constant $\alpha_s$ (the value of $\alpha_s$ is around~0.3 in our simulations).
This action has excellent rotational invariance properties in the static quark potential even on coarse lattices~\cite{Alford:1995hw}. 
The scale for the pure gauge L\"uscher-Weisz action has been set in an earlier study~\cite{Gattringer:2001jf}.

\begin{figure}[t]
  \centering
  \includegraphics[width=0.5\columnwidth,trim= 0 0 0 -1.5cm]{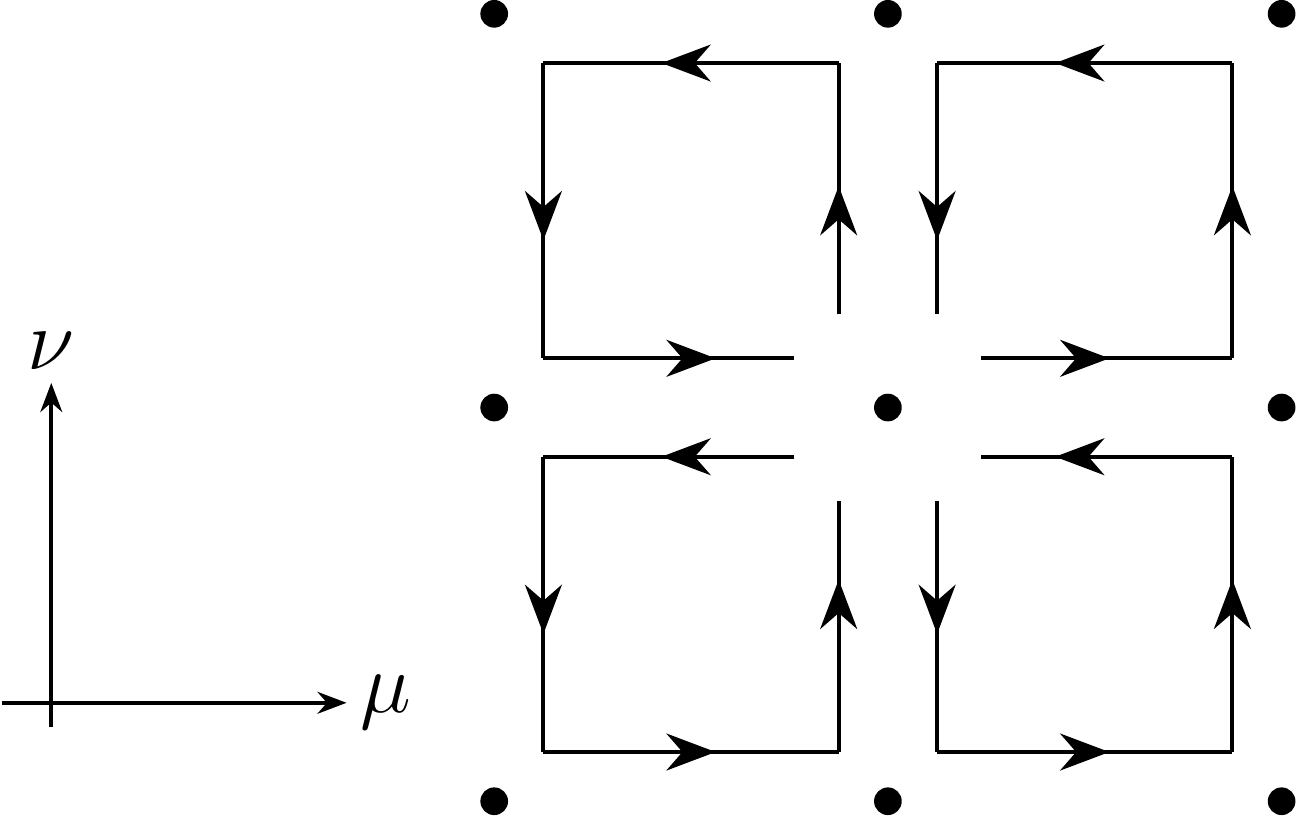} 
  \caption{The clover field in Eq.~\protect\ref{eq:clover} constructed from four plaquettes in the $\mu\nu$ plane.
  }
  \label{fig:clover}
\end{figure}

The fermion action is based on the clover discretization.
The Dirac operator can be written as 
\beq
D_\text{clover}=D_\text{W} - \frac12\kappa \, c_{\rm SW}  \sum_{\mu\nu}\sigma_{\mu\nu} F_{\mu\nu}
\eeq
where $D_\text{W}$ is the Wilson term 
\beqs
D_\text{W}[V](x,y) =\delta_{x,y} -\kappa &\sum_{\mu}\big[(r-\gamma_\mu)V_\mu(x) \delta_{x+\mu,y} \\
&+ (r+\gamma_\mu)V_\mu(y)^\dagger \delta_{x,y+\mu} \big]\,.
\eeqs
We work with $r=1$.
For the clover term, $\sigma_{\mu\nu} = \frac{1}{2}[\gamma_\mu,\gamma_\nu]$
and the clover field $F$ is constructed from the anti-hermitian and traceless part of the average of four plaquetes,
\beq
F_{\mu\nu} =  \frac{1}{8}[Q_{\mu\nu}-Q^\dagger_{\mu\nu}],
\label{eq:Fmunu}
\eeq 
where 
\beq
Q_{\mu\nu} =
 V_{\mu\nu-\mu-\nu} + V_{\nu-\mu-\nu\mu} +  V_{-\mu-\nu\mu\nu} + V_{-\nu\mu\nu-\mu},
\label{eq:clover}
\eeq
is represented pictorially in Fig.~\ref{fig:clover}.

The clover coefficient is set to its tree-level value $c_{\rm SW}=1.0$ since we are using nHYP smeared links in the fermionic kernel, and non-perturbative determination of the clover coefficient found the corrections to be insignificant~\cite{Hoffmann:2007nm}.

We now turn to the smearing function:  nHYP smearing~\cite{Hasenfratz:2007rf} refers to {\it normalized hypercubic smearing} which 
involves three consecutive levels of APE-style smearing~\cite{Albanese:1987ds},
\beqs
V_{n,\mu} = {\cal P} \Bigg[&(1-\alpha_1)U_{n,\mu} \\
&+ \frac{\alpha_1}{6}\sum_{\pm\nu\neq\mu}
\tilde{V}_{n,\nu;\mu} \tilde{V}_{n+\hat{\nu},\mu;\nu} \tilde{V}^\dagger_{n+\hat{\mu},\nu;\mu}\Bigg],\\
\tilde{V}_{n,\mu;\nu} = {\cal P}\Bigg[&(1-\alpha_2)U_{n,\mu} \\
&+\frac{\alpha_2}{4}\sum_{\pm\rho\neq \nu,\mu}
\bar{V}_{n,\rho;\nu\mu} \bar{V}_{n+\hat{\rho},\mu;\rho\nu} \bar{V}^\dagger_{n+\hat{\mu},\rho;\nu\mu}\Bigg],\\
\bar{V}_{n,\mu;\nu\rho} = {\cal P}\Bigg[&(1-\alpha_3)U_{n,\mu} \\
&+\frac{\alpha_3}{2}\sum_{\pm\eta\neq\rho,\nu,\mu}
U_{n,\eta} U_{n+\hat{\eta},\mu} U^\dagger_{n+\hat{\mu},\eta}\Bigg]\,.
\eeqs
Here the conventional smearing notation using $U_{n,\mu}$ to represent $U_\mu(x)$ is adopted.
The intermediate fields $\tilde{V}$ and $\bar{V}$ are constructed such that the contributions to $V_{n,\mu}$ are restricted to the thin links $U$ appearing as edges in the hybercubes attached to the link $V_{n,\mu}$. The indices after the semicolons indicate the directions excluded from the sums.
The operator ${\cal P}$ is a projection to $U(3)$,
\beq
{\cal P} A \equiv A(A^\dagger A)^{-1/2}.
\eeq
The projection is non-singular by construction since $(A^\dagger A)^{-1/2}$ is Hermitian and positive definite.
The three $U(3)$ projections render the nHYP smeared configurations very smooth, while keeping the smearing within a hypercube ensures that even short distance properties of the configurations are minimally distorted.
Consequently, the fermion action is ultra local.
Unlike HYP links which project to $SU(3)$ rather than $U(3)$,  
the nHYP links are differentiable, which makes them amenable to HMC algorithms. 
We use the standard HYP values
$\alpha_1=0.75$, $\alpha_2=0.6$, and $\alpha_3=0.3$
that have been tuned to minimize the plaquette fluctuations~\cite{Hasenfratz:2001hp}.

The idea of using smeared links in the fermion action to improve its scaling properties has been validated in a number of studies. In Ref.~\cite{Alexandru:2004ge} it is shown that staggered fermions with HYP smearing has very good scaling properties, even better than other improved actions. In Ref.~\cite{Hasenfratz:2007rf} and Ref.~\cite{Hoffmann:2007nm}, the same nHYP fermion plus L\"uscher-Weisz gauge action is explored on small lattices. The clover coefficient $c_{\rm SW}$ is computed non-perturbatively and found to be close to the tree-level value of 1.0 with small corrections.
In Ref.~\cite{D_rr_2009}, tree-level clover fermion with stout links and tree-level Symanzink gauge action are used. A wide scaling region up to lattice spacing of 0.16 fm is found.
Stout smearing~\cite{Morningstar_2004} is an alternative projection method that is also differentiable. It can be used to replace the projection ${\cal P}$ discussed above, leading to the so-called HEX smearing. A two-level HEX smeared action is used
in Refs.~\cite{D_rr_2009,D_rr_2011} to study continuum limit for hadron masses and quark mass renormalization.

%
%

%
\begin{table*}[t]
  \centering
  \caption{
  Ensemble parameters for the dynamical simulations on the $24^3 \times 48$ lattice used in this study. Also presented are Wilson flow parameters $w_0/a$ and $t_0/a^2$ defined in Eq.~\protect\ref{eq:t0w0}, pion mass $am_\pi$, lattice spacing in fm determined from $w_0^{\rm chiral}$, finite volume corrections for the pion mass in Eq.~\protect\ref{eq:R} from NNLO $\chi$PT,  and input and measured tadpole factor $u_0$.}
  \label{tab:latParams}
  \def\arraystretch{1.35}
  \begin{tabular*}{\textwidth}{@{\extracolsep{\fill}} L L L L L L c L L}
    \hline
    \multicolumn{1}{c}{$\beta$} & \multicolumn{1}{c}{$\kappa$} &
    \multicolumn{1}{c}{$w_0/a$} & \multicolumn{1}{c}{$t_0/a^2$} &
    \multicolumn{1}{c}{$am_\pi$} &
    \multicolumn{1}{c}{$a[\fm]$} &
    \multicolumn{1}{c}{$|R_{m_\pi}|(\%)$} &
    \multicolumn{1}{c}{$u_0^{\rm input}$} & \multicolumn{1}{c}{$u_0^{\rm measured}$} \\[1pt]
    \hline
    7.2 & 0.1276 & 1.4494(29) & 1.6859(40) & 0.24723(52) & 0.11481(87) & 0.05 & 0.8666 & 0.866225(2) \\
    & 0.1278 & 1.4699(26) & 1.7185(33) & 0.20392(51) & 0.11533(87) & 0.11 & 0.8666 & 0.866271(1) \\
    & 0.1280 & 1.5068(32) & 1.7719(43) & 0.15097(74) & 0.11469(88) & 0.39 & 0.8666 & 0.866319(1) \\
    \hline
    7.3 & 0.1273 & 1.6173(34) & 2.1134(51) & 0.21555(59) & 0.10323(79) & 0.11 & 0.8697 & 0.870022(1) \\
    & 0.1275 & 1.6681(46) & 2.1914(65) & 0.16579(68) & 0.10231(80) & 0.34 & 0.8697 & 0.870061(1) \\
    & 0.1277 & 1.7039(50) & 2.2498(67) & 0.10758(91) & 0.10238(81) & 1.50 & 0.8697 & 0.870098(1) \\
    \hline
    7.4 & 0.1268 & 1.7900(32) & 2.6253(53) & 0.23203(54) & 0.09111(69) & 0.11 & 0.8735 & 0.873693(1) \\
    & 0.1270 & 1.8312(43) & 2.7022(67) & 0.1939(11) & 0.09097(70) & 0.23 & 0.8735 & 0.873711(1) \\
    & 0.1272 & 1.8747(77) & 2.778(12) & 0.1506(15) & 0.09088(76) & 0.57 & 0.8735 & 0.873736(1) \\
    \hline
    7.5 & 0.1266 & 1.9868(67) & 3.260(13) & 0.21221(96) & 0.08190(67) & 0.22 & 0.8769 & 0.877045(1) \\
    & 0.1268 & 2.0465(91) & 3.376(17) & 0.1732(13) & 0.08142(72) & 0.48 & 0.8769 & 0.877068(1) \\
    & 0.1270 & 2.0936(74) & 3.464(14) & 0.1261(12) & 0.08179(67) & 1.29 & 0.8769 & 0.877083(1) \\
    \hline
    7.6 & 0.1260 & 2.1320(69) & 3.838(14) & 0.2576(11) & 0.07266(61) & 0.02 & 0.8800 & 0.880110(1) \\
    & 0.1262 & 2.1653(76) & 3.925(16) & 0.23487(64) & 0.07268(61) & 0.24 & 0.8800 & 0.880122(1) \\
    & 0.1264 & 2.207(10) & 4.011(19) & 0.2019(15) & 0.07305(66) & 0.48 & 0.8800 & 0.880136(1) \\
    & 0.1266 & 2.242(10) & 4.112(22) & 0.1690(12) & 0.07371(66) & 0.79 & 0.8800 & 0.880155(1) \\
    & 0.1267 & 2.2794(97) & 4.186(20) & 0.1493(17) & 0.07345(65) & 1.17 & 0.8800 & 0.880166(1) \\
    \hline
    7.7 & 0.1256 & 2.245(10) & 4.422(26) & 0.2843(12) & 0.06680(61) & 0.00 & 0.8831 & 0.883014(1) \\
    & 0.1258 & 2.321(17) & 4.612(44) & 0.25879(99) & 0.06546(74) & 0.00 & 0.8831 & 0.883026(1) \\
    & 0.1260 & 2.400(16) & 4.827(38) & 0.23300(90) & 0.06430(69) & 0.00 & 0.8831 & 0.883041(1) \\
    & 0.1263 & 2.480(17) & 5.030(41) & 0.1827(19) & 0.06483(73) & 0.46 & 0.8831 & 0.883064(1) \\
    & 0.1265 & 2.476(16) & 5.032(36) & 0.1519(17) & 0.06679(68) & 1.09 & 0.8831 & 0.883067(1) \\
    \hline
  \end{tabular*}
\end{table*}
%

\section{\label{sec:lsp}Scale setting}

\subsection{Simulation parameters}

We carried out dynamical simulations on $24^3 \times 48$ lattices with 6 values of $\beta$ and up to 5 values of $\kappa$ for each $\beta$, totaling 22 different ensembles.
The gauge fields are updated using the
HMC algorithm~\cite{Duane:1987de}. We employ the Hasenbush-Jansen multi-mass method~\cite{Hasenbusch:2001ne} to
speed up the calculation: by using two masses and adjusting the integration time steps for gauge and fermion force updates to keep the contribution balanced between the terms.
We set the trajectory length to $1$ and varied the number of integration steps to keep the acceptance rate above 90\%.
For each ensemble we
generated 500 configurations using 2500 trajectories and saving every fifth trajectory.
We then dropped the first 100 configurations for thermalization and kept the last 400 for calculations.

The list of all the ensembles is given in Table~\ref{tab:latParams} along with other quantities determined in this work. The values for $\beta$ were chosen to scan finely the $0.05\fm\lesssim a \lesssim 0.1\fm$ range. Since our simulations were run on a fixed sized lattice,
to keep the finite volume errors under control, we used $\kappa$ values such that
$0.15 \lesssim a m_\pi \lesssim 0.3$. For each ensemble we compute the pion mass using two point correlation functions with several sources per configuration, evenly spread out in the Euclidean time direction. The quark propagators were computed using our optimized GPU inverters~\cite{Alexandru:2011ee}. We perform single exponential fits
to extract the pion mass. The fit range for the pion correlators were determined by looking for plateaus in the effective mass plots for the correlators.

Since the observables are determined to percent, or better, precision level, we ensure that the systematic errors remain below percent level. In particular, we must assess finite volume effects for the pion mass. The finite volume effects for $t_0$ and $w_0$, the other observables used in this study, are smaller~\cite{Borsanyi:2012zs}.

Finite volume corrections to the pion mass can be related to the pion forward scattering
amplitude~\cite{Luscher:1985dn}.
Alternatively, such corrections can be computed using chiral perturbation theory in finite
volume~\cite{Gasser:1986vb,Gasser:1987ah,Gasser:1987zq}.
The two approaches were compared in Refs.~\cite{Colangelo:2002hy,Colangelo:2003hf}, where
the subleading effect in both approaches, the next leading exponential correction for L\"uscher's
approach and NLO effects for the \chipt approach, were found to produce large corrections.
However, as the \chipt NNLO effects were found to be small, finite-volume \chipt is expected
to be reliable.
To improve the convergence of L\"uscher's method~\cite{Luscher:1985dn}, a resummation method
was proposed in Ref.~\cite{Colangelo:2005gd} for finite-volume correction on the pion mass, 
\beqs 
&R_{m_\pi}=\frac{m_\pi(L)-m_\pi}{m_\pi} 
= -\frac{1}{32\pi^2 m_\pi L} \sum_{n=1}^\infty \frac{m(n)}{ \sqrt{n}} \\ 
& \times \int_{-\infty}^\infty d\tilde{y}\, {\cal F}_\pi(i\tilde{y}) \,e^{-\sqrt{n(1+\tilde{y}^2)}\;m_\pi L} + O(e^{-m_\pi L}),
\label{eq:R}
\eeqs
where $m(n)$ are multiplicities whose values can be found in Table 1 of Ref.~\cite{Colangelo:2005gd}.
The resummed correction shows very good agreement with the
finite volume \chipt expansion: when using the LO(NLO) \chipt expression in the forward
amplitude ${\cal F}_\pi(i\tilde{y})$, the results agree with 1-loop(2-loop)
finite-volume \chipt expansion. The resummed method was checked by direct 2-loop
calculation of the finite-volume \chipt and found to be accurate for
$m_\pi L \gtrsim 2$~\cite{Colangelo:2006mp,Bijnens:2014dea}. 
Our values for $m_\pi L$ from Table~\ref{tab:latParams} lie between 2.6 to 6.7.

We use the \chipt results to estimate the expected finite volume corrections: 
we interpolated the results from Table 3 of Ref.~\cite{Colangelo:2005gd} to get the magnitude of these corrections for our ensembles. The resulting estimates are summarized in Table~\ref{tab:latParams}. For our ensembles these corrections are at the level of 1\% or below.

\subsection{Lattice spacing methodology}

There are several well-established methods used to determine the lattice spacing: the Sommer parameter~\cite{Sommer:1993ce}, hadron masses~\cite{Gattringer:2010zz}, etc.
While scale setting using hadron masses is conceptually straightforward, such quantities are often not
determined with very high precision, and are computationally quite expensive. Additionally, systematic
uncertainties associated with extracting hadron masses, such as excited state contamination
and finite size effects, are not always easy to estimate.
The Sommer scale $r_0$ is based on the calculation of the static
potential from the gauge fields, which does not require the computation of expensive quark propagators. 
However, care must be taken to select the fit form and ranges.
In addition to using observables which may be determined with good accuracy at minimal computational cost,
it is often advantageous to use scales with a mild quark mass dependence~\cite{Sommer:2014mea}. 
The Wilson flow method proposed by L\"{u}scher~\cite{Luscher:2010iy} satisfies all these requirements.
This uses new parameters $t_0$ and $w_0$~\cite{Bruno:2013gha,Borsanyi:2012zs}, defined below, to set the scale.

The Wilson flow is essentially a smearing of the original
gauge fields $U_\mu(x)$ controlled by a 
{\em flow time} parameter $t$, not to be confused with $x_4$, the Euclidean time coordinate.  
This method has the advantage that it is very straightforward to implement.
The Wilson flow for gauge fields $U_\mu(x)$ is defined by
the first-order differential equation
\beqs
  \partial_t V_\mu(x,t) &= -\frac{6}{\beta} \{\partial_{x,\mu}S_W[V_\mu(x,t)]\} V_\mu(x,t),\\
  V_\mu(x,t)\big\rvert_{t=0} &= U_\mu(x),
\eeqs
where $S_W$ is the standard Wilson gauge action (the first term in Eq.~\ref{eq:LW}) evaluated using the smeared links $V_\mu(x,t)$. 
The smearing radius is proportional to $\sqrt{t}$.
We numerically integrate this equation for a flow time $t_0$ such that the smearing radius achieves a particular {\em physical} value.
To determine this point we monitor the energy density given by
\beq
\av{E(t)} \equiv -\frac{1}{2}\av{\Tr\left\{F_{\mu\nu}(t)F_{\mu\nu}(t)\right\}}.
\eeq
For the gluon field strength tensor $F_{\mu\nu}$ we use the clover discretization in Eq.~\ref{eq:Fmunu} constructed from the smeared fields $V_\mu(x,t)$.
The parameters $t_0$ and $w_0$ are defined implicitly by the dimensionless, renormalized quantities
\beqs
\label{eq:t0w0}
    t^2\av{E(t)}&\big\rvert_{t=t_0}\; = 0.3, \\
    t\dd{}{t}\big(t^2\av{E(t)}\big)&\bigg\rvert_{t=w_0^2} = 0.3.
\eeqs
An advantage of using $t_0$ and $w_0$ for scale setting is that, despite their
relatively large auto-correlations, their stochastic errors are small and can be
computed precisely and are computationally inexpensive.
As $t_0$ and $w_0$ are not experimentally measurable quantities, in order to assign physical
units we must compare to other lattice calculations in which a physical quantity has
been used to set the scale (e.g. using hadron mass ratios). Hence, these Wilson flow observables
offer an efficient way of determining the relative scale/lattice spacing between many simulations
with differing gauge couplings. For $N_f=2$ we refer to Ref.~\cite{Bruno:2013gha} where
$t_0$ and $w_0$ are reported in physical units at the chiral point.
For a recent application of Wilson flow to set the scale for two-color QCD, see Ref.~\cite{Iida:2020emi}.

\subsection{\label{sec:results}Lattice spacing results}

We measure $w_0/a$ from Eq.~\ref{eq:t0w0} on multiple
ensembles with the same gauge coupling $\beta$ and extrapolate to the chiral point
($m_\pi = 0$). We fix the scale by the value of $w_0$ at the chiral point $w_0^{\rm chiral}$. 
We then use these measurements to find a parameterization of the lattice spacing $w_0^{\rm chiral}/a$
as a function of $\beta$.
We are able to determine $w_0/a$ with very high accuracy at a few parts per-mille level.

To convert the lattice spacing results to physical units we use
the $N_f=2$ value of $w_0^{\rm chiral}=0.1776(13)\fm$ from
Ref.~\cite{Bruno:2013gha}. This value is extracted by computing
the kaon decay constant in $N_f=2$ simulations~\cite{Fritzsch:2012wq}.
Note that in this work, when we report the lattice spacing results in physical units, the bulk of the uncertainty in the lattice spacing comes from the uncertainty in $w_0^{\rm chiral}$.

The extrapolation of $w_0/a$ to the chiral point requires a detailed discussion. Following Ref.~\cite{Bruno:2013gha}, we compute $w_0/a$ as a function of the variable $y=(m_\pi w_0)^2$ using the ensemble data\footnote{Notice that our definition of $y$ here is different from the definition in Ref.~\cite{Bruno:2013gha} which is $y=m_\pi^2 t_0$.}. If the values of $a$ are properly determined we can compute then $w_0$ as a function of $y$ and these data points are expected
to lie on a universal curve, up to discretization errors.
In Ref.~\cite{Bruno:2013gha} this curve was found to be well
described by a line, but our data-points have a wider range of $y$
values and we see a deviation from the linear behavior. 
To account for the non-linear behavior, we perform a quadratic fit to $w_0/w_0^{\rm chiral}$ as a function of $y$
for each $\beta$ and use the resulting function to interpolate $w_0/a$ as a function of $y$.

To obtain the universal curve for $w_0(y)$ we use the following procedure. For each $\beta$ we introduce a fit parameter $(a/w_0^{\rm chiral})_\beta$. Using  these parameters we construct a curve $[w_0/w_0^{\rm chiral}](\beta)$ and fit it to a quadratic function constrained to pass through $1$ for $y=0$
\begin{align}
  \begin{split}
    w_0/w_0^{\rm chiral} &\equiv \left(\frac{w_0}{a}\right) \times \left(a/w_0^{\rm chiral}\right)_{\beta}, \\
    [w_0/w_0^{\rm chiral}](y) &= (1 - c_1 y + c_2 y^2).
    \label{eq:w0fit}
  \end{split}
\end{align}
There are in total 8 fit parameters, $c_1$, $c_2$, and six different $(a/w_0^{\rm chiral})_\beta$.
In order to estimate the uncertainty associated with these fit parameters, standard $\chi^2$ fitting does not work since we have statistical uncertainty associated with both $w_0/a$ and $y$. The alternative approach that we use is a variant of jackknife  resampling. We bin the configurations in steps of $20$ to take into account the auto-correlations, which are particularly large for $w_0/a$ measurements. For each jackknife sample we compute the values of $w_0/a$ and $y$.
For each sample we perform a $\chi^2$ minimization with $\chi^2$ defined as
\beqs
&\chi^2 = \sum_{\beta,\kappa} 
\left(
\frac{(w_0/a)_{\beta,\kappa} (a/w_0^{\rm chiral})_{\beta} - f(y_{\beta,\kappa})}
{\sigma_{(w_0/a)_{\beta,\kappa}} \times (a/w_0^{\rm chiral})_{\beta}}
\right)^2 \\
&\text{with} \quad f(y) = 1 - c_{1}\,y + c_{2}\,{y}^2\,. \\
\eeqs

The error estimate $\sigma_{(w_0/a)}$ is the same for all samples and it is estimated from the fluctuations of this observable over the samples. On the other hand both $(w_0/a)$ and $y$ change from sample to sample. This minimization produces for each sample eight fit parameters corresponding to $c_{1}$, $c_{2}$, and six values of $(a/w_0^{\rm chiral})_\beta$. We use these values to estimate fit parameters and their uncertainties. The results are presented in Table~\ref{tab:params}.

\begin{figure}[t]
  \centering
  \includegraphics[width=\columnwidth,trim= 0 0 0 -0.5cm]{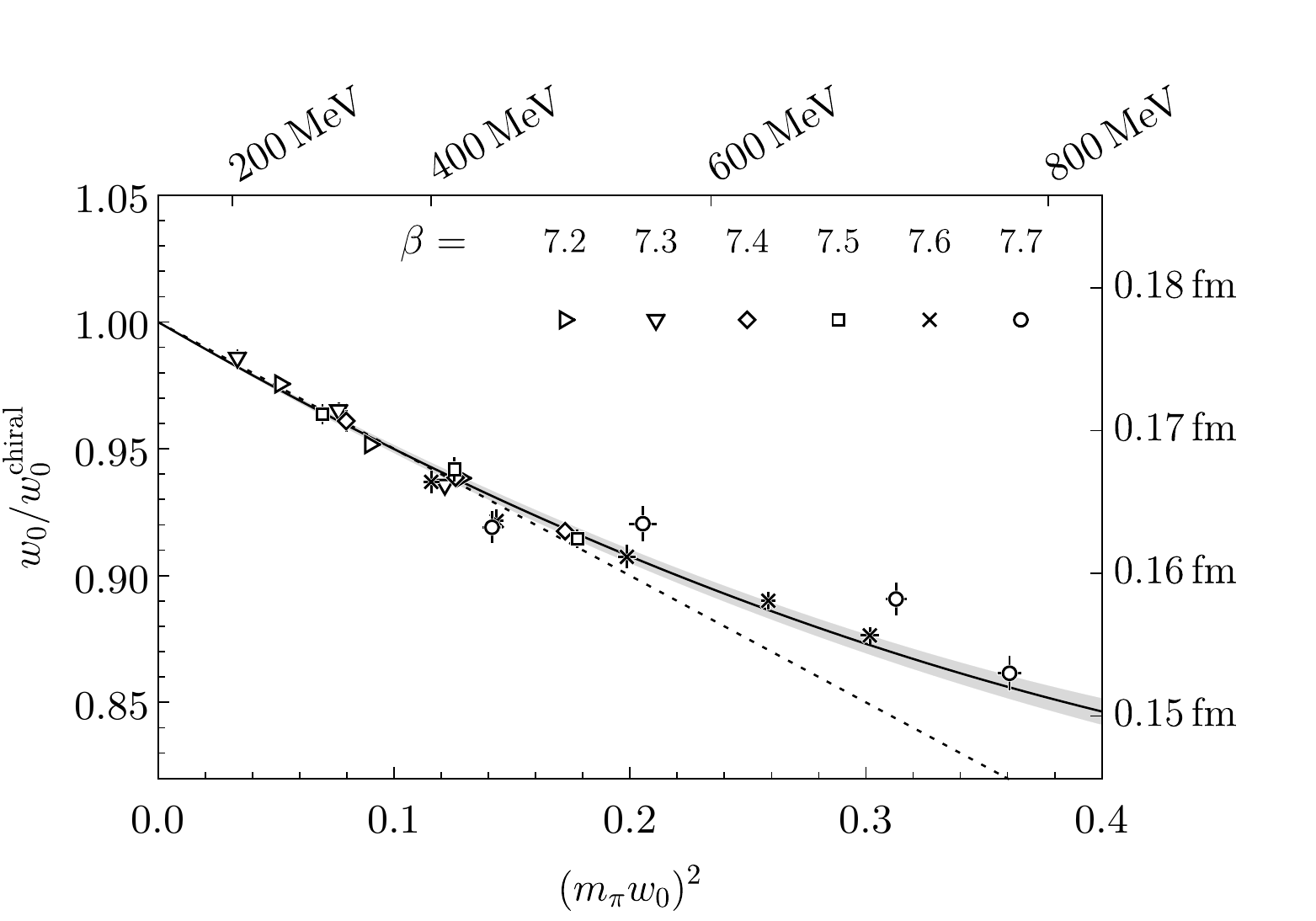}
  \caption{Chiral extrapolation for the $w_0$ parameter as a function of $y=(m_{\pi} w_0)^2$. 
  The solid line is the quadratic fit
    describing our universal curve and the dashed line is the linear universal curve inferred
    from Ref.~\protect\cite{Bruno:2013gha}. To guide the reader we also display on the top the physical pion
    mass corresponding to $y$ and on the right the $w_0$ in physical units. These values are computed
    using $w_0^{\rm chiral}=0.1776\fm$.
  }
  \label{fig:w0universal}
\end{figure}

\begin{table}[b]
  \def\arraystretch{1.35}
  \begin{tabular*}{0.8\columnwidth}{@{\extracolsep{\fill}}ccrc}
  \toprule
  $\beta$ & $a/w_0^{\rm chiral}$ &\phantom{andrei}& $a[\fm]$ \\
  \midrule
  $7.2$ & $0.6474(12)$ && $0.11499(21)$ \\
  $7.3$ & $0.5786(13)$ && $0.10276(24)$ \\
  $7.4$ & $0.5126(15)$ && $0.09103(26)$ \\
  $7.5$ & $0.4603(17)$ && $0.08174(29)$ \\
  $7.6$ & $0.4111(14)$ && $0.07301(26)$ \\
  $7.7$ & $0.3711(21)$ && $0.06591(37)$ \\
  \midrule
  $c_1$ & $0.541(19)$ &  $c_2$ & $0.390(45)$ \\
\bottomrule
  \end{tabular*}
    \caption{Fit parameters from the fit defined in Eq.~\protect\ref{eq:w0fit}. The last column displays
   the fit results for lattice spacing converted to physical
   units using $w_0^{\rm chiral}=0.1776\fm$.}
  \label{tab:params}
\end{table}

\def\wch{w_0^{\rm chiral}}

\begin{figure}[t]
  \centering
  \includegraphics[width=0.9\columnwidth,trim= 0 0 0 -0.5cm]{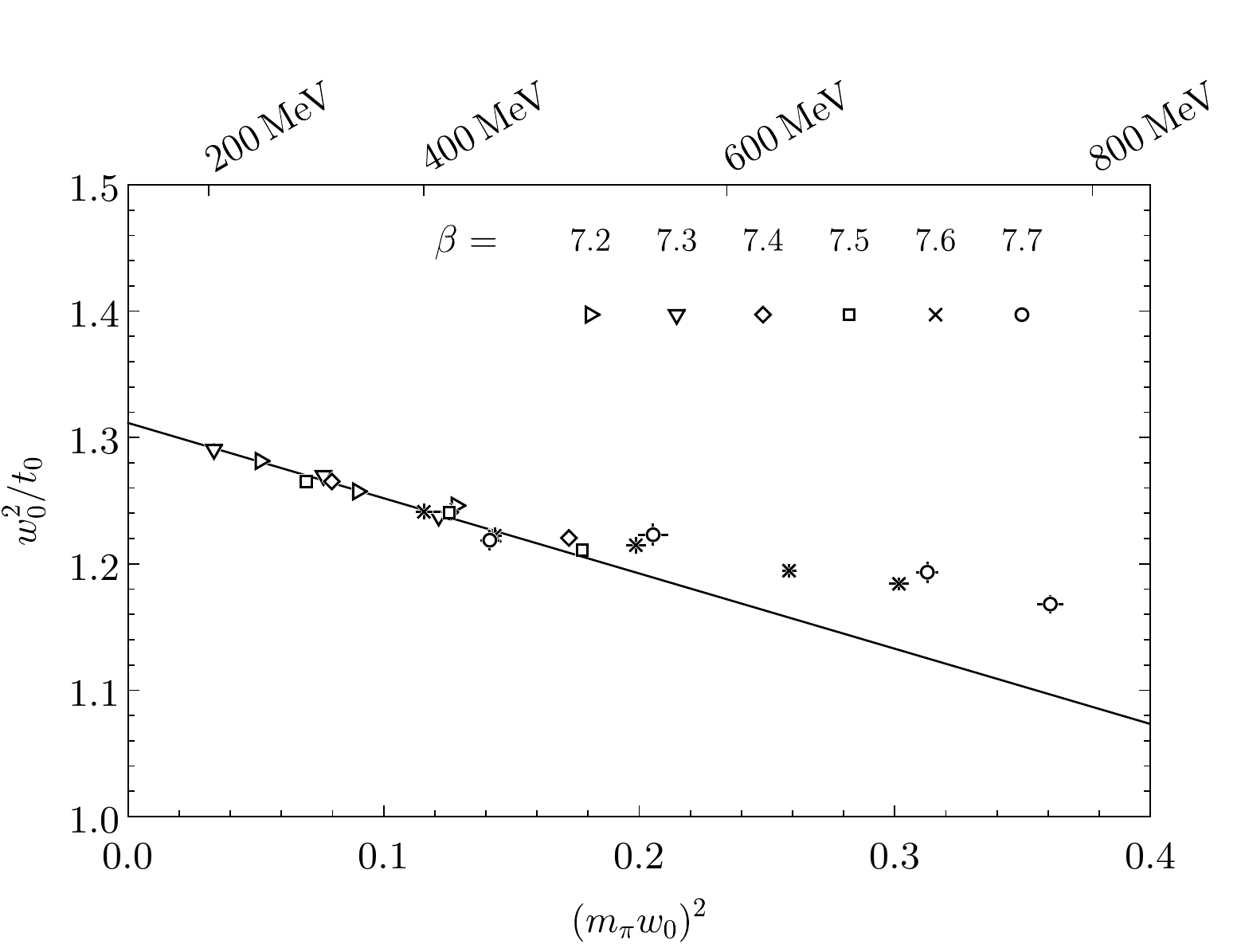}
  \caption{Similar to Fig.~\protect\ref{fig:w0universal}, but for the dimensionless ratio $w_0^2/t_0$. 
  The solid line is a linear fit for the values with $y\leq 0.1$.
  }
  \label{fig:w0sqovt0}
\end{figure}

The results of the fit are presented in Fig.~\ref{fig:w0universal}. The measurements for $w_0/a$ from Table~\ref{tab:latParams} are rescaled using the $(a/\wch)_\beta$ fit parameters from Table~\ref{tab:params} and the solid line represents the chiral extrapolation using the $c_{1,2}$ fit results. The dotted line represents the linear chiral extrapolation presented in Ref.~\cite{Bruno:2013gha}. We convert the results in Fig. 4 of Ref.~\cite{Bruno:2013gha},  $t_0/(t_0)_\text{ref}\approx 1.08- t_0 m_\pi^2$, to straight line in our extrapolation of $w_0$ as a function of $(m_\pi w_0)^2$. For this we calculate $w_0^2/t_0$ as a function of $y$; the results are plotted in Fig.~\ref{fig:w0sqovt0}. For small values of $y$ this ratio can be fit with a straight line and we find the ratio of $w_0^2/t_0$ at the chiral point to the value at $m_\pi=390\MeV$, the reference point used in Ref.~\cite{Bruno:2013gha}, to be $1.31/1.26$. Since the ratio of $t_0$ between these points is $1.08$, we find that the ratio of $w_0$ is $1.06$, which leads to a slope of $0.50$ since at the reference mass $y=0.11$.

Using the functional form $w_0(y)/\wch=1-c_1 y+c_2 y^2$ we can extrapolate the $w_0/a$ measurements in Table~\ref{tab:latParams} to the chiral limit and extract the lattice spacing using the physical value for $\wch$. This procedure is used to compute the lattice spacing column in that table and we see that it produces consistent results for the same $\beta$ but different $\kappa$ values.
Recall that the error in $w_0/a$ is on the order of a few parts per thousand, so that
the overwhelming source of statistical error in the lattice spacing determination comes
from $w_0^{\rm chiral}$. More accurate determinations of $w_0^{\rm chiral}$ in physical units
will therefore have a significant impact on the accuracy of the lattice spacing determination.

\section{\label{sec:param}Parameterizations}
The results so far are at selected values of $\beta$ and $\kappa$. In this section we perform various interpolations parameterized by smooth functions, to facilitate the usage of the action at other desired lattice spacings and pion masses.

\subsection{Lattice spacing}
\begin{figure}[t]
  \centering
  \includegraphics[width=1.0\columnwidth]{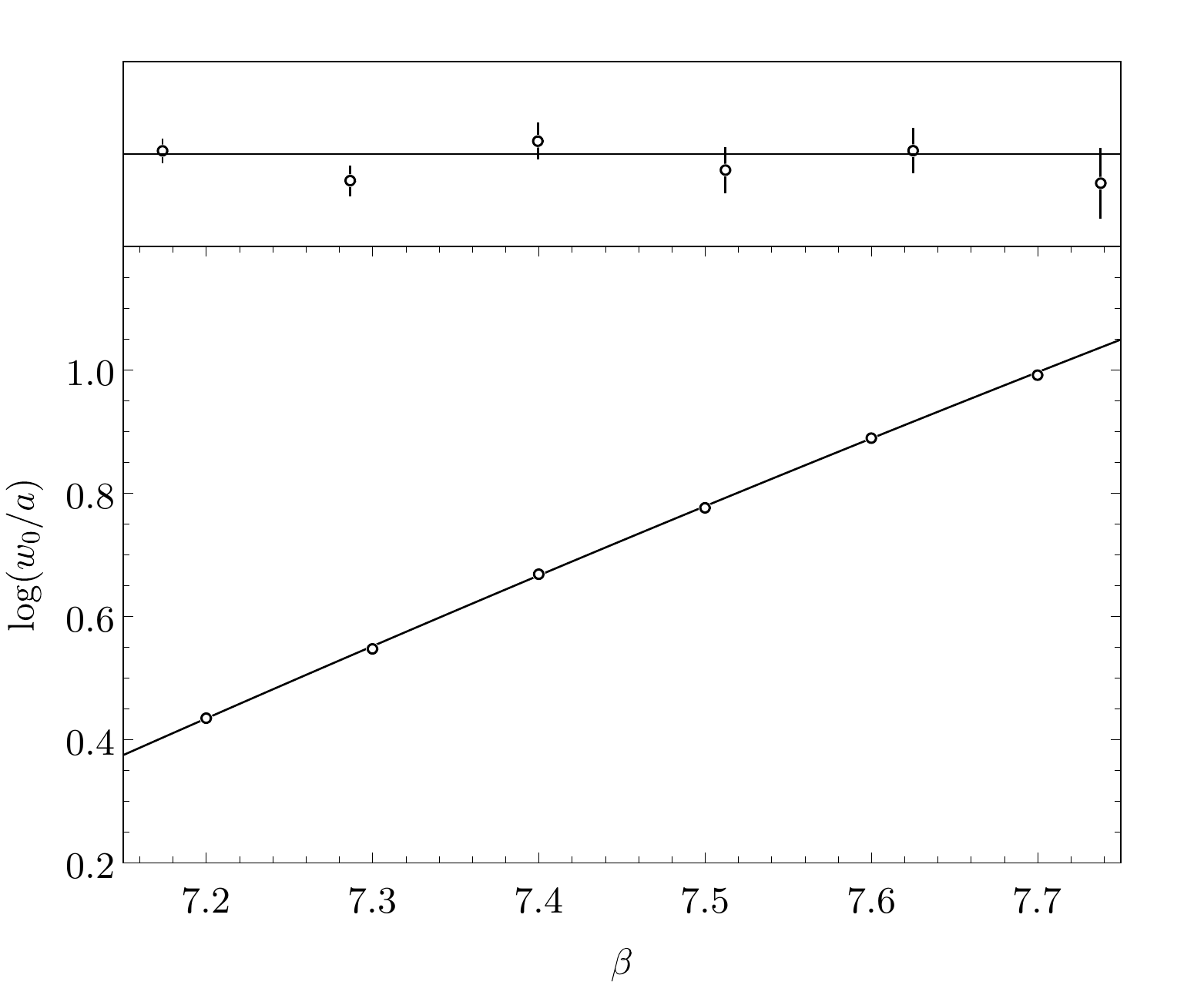}
  \caption{Lattice spacing as a function of $\beta$. Here $w_0$
  is defined in the chiral limit. The parametrization in Eq.~\ref{eq:a_vs_beta} describes very well the measurements.
  The plot above represents the differences between the 
  parametrization and the measured values.
  }
  \label{fig:a_vs_beta}
\end{figure}

Using the values for $w_0/a_\beta$ coming from the global fit described in Section.~\ref{sec:lsp}, we can perform a quadratic fit to find a parameterization of $a$ as a function of $\beta$. The result is 
\beqs
\log{(w_0/a)} = 0.5514 + 1.1600\Delta\beta - 0.1217(\Delta\beta)^2 \,,
    \label{eq:a_vs_beta}
\eeqs
where $\Delta\beta\equiv \beta-7.3$.
We use $\beta=7.3$ as the expansion point since the lattice spacing is close to $0.1 \fm$ here. 
 This fit is illustrated in Fig.~\ref{fig:a_vs_beta}. The error bars are plotted in the figure, but due to the small errors on the $a_\beta$ values, they are not visible. 
 
Using a physical determination for $w_0$, this parameterization can be used to obtain the $\beta$ value corresponding to a desired lattice spacing $a$ in physical units. The function is fairly smooth and we expect that it works reasonably well even outside the range of $\beta$ used in our simulations.

\subsection{Tadpole factor}

The tadpole factor $u_0$ is an important input parameter that guarantees the 
improved properties of the L\"uscher-Weisz gauge action.
Prior to generating our ensembles on the $24^3\times 48$ lattice, we did a sensitivity study of pion mass on the input value of $u_0$. To this end we generated two $16^3\times 32$ ensembles at $\beta = 7.3$ and $\kappa = 0.1275$, each with 800 configurations (4000 thermalized trajectories separated in steps of five). 
We used two values for $u_0$, $0.868$ and $0.870$, and found that the pion mass changed from $a m_\pi=0.188(2)$ to $0.179(2)$, a deviation of 5\%. We conclude that changes of $u_0$ in the fourth significant digit should produce sub-percent level shifts for the pion mass.
Both the input value and measured value for $u_0$ are given in Table~\ref{tab:latParams}. The measured value of $u_0$ is very close to the input value.
The variation of $u_0$ in the $\kappa$ values used in our simulations is very small and we decided to use the same input $u_0$ for all values of $\kappa$. 

\begin{figure}[t]
  \centering
  \includegraphics[width=1.0\columnwidth]{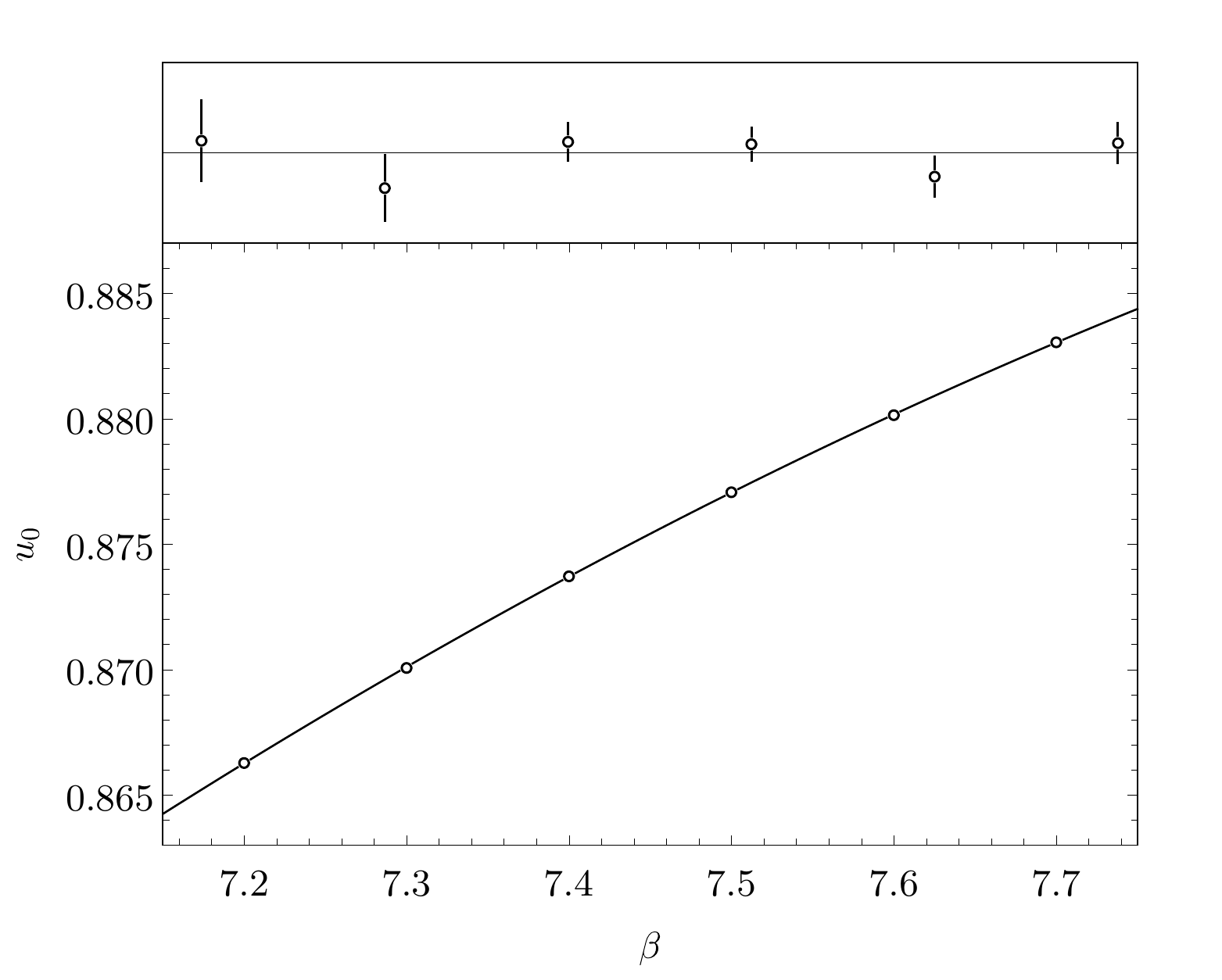}
  \caption{Tadpole improvement factor $u_0$ as a function $\beta$. 
  The inset at the top shows the difference between the parameterization in Eq.~\ref{eq:u0interp} and the
  measured values. The error bars correspond to the standard deviation of the $u_0$ values measured at different quark mass values.
  }
  \label{fig:u0_vs_beta}
\end{figure}

To interpolate $u_0$ as a function of $\beta$, we use the average measured values of $u_0$ for the ensembles with the same $\beta$ to perform a quadratic fit. The result is a smooth function given by
\beq
    u_0(\beta) = 0.87010 + 0.03721 \Delta\beta 
    -0.01223(\Delta\beta)^2\,,
    \label{eq:u0interp}
\eeq
and displayed in Fig.~\ref{fig:u0_vs_beta}.

This function can be used in conjunction with the function in Eq.~\ref{eq:a_vs_beta} to pick the input~$\beta$ and~$u_0$ values corresponding to a desired lattice spacing~$a$.

\subsection{Pion mass}
\begin{figure}[b]

  \centering
  \includegraphics[width=0.99\columnwidth,trim=0 0 0.4cm 0.5cm]{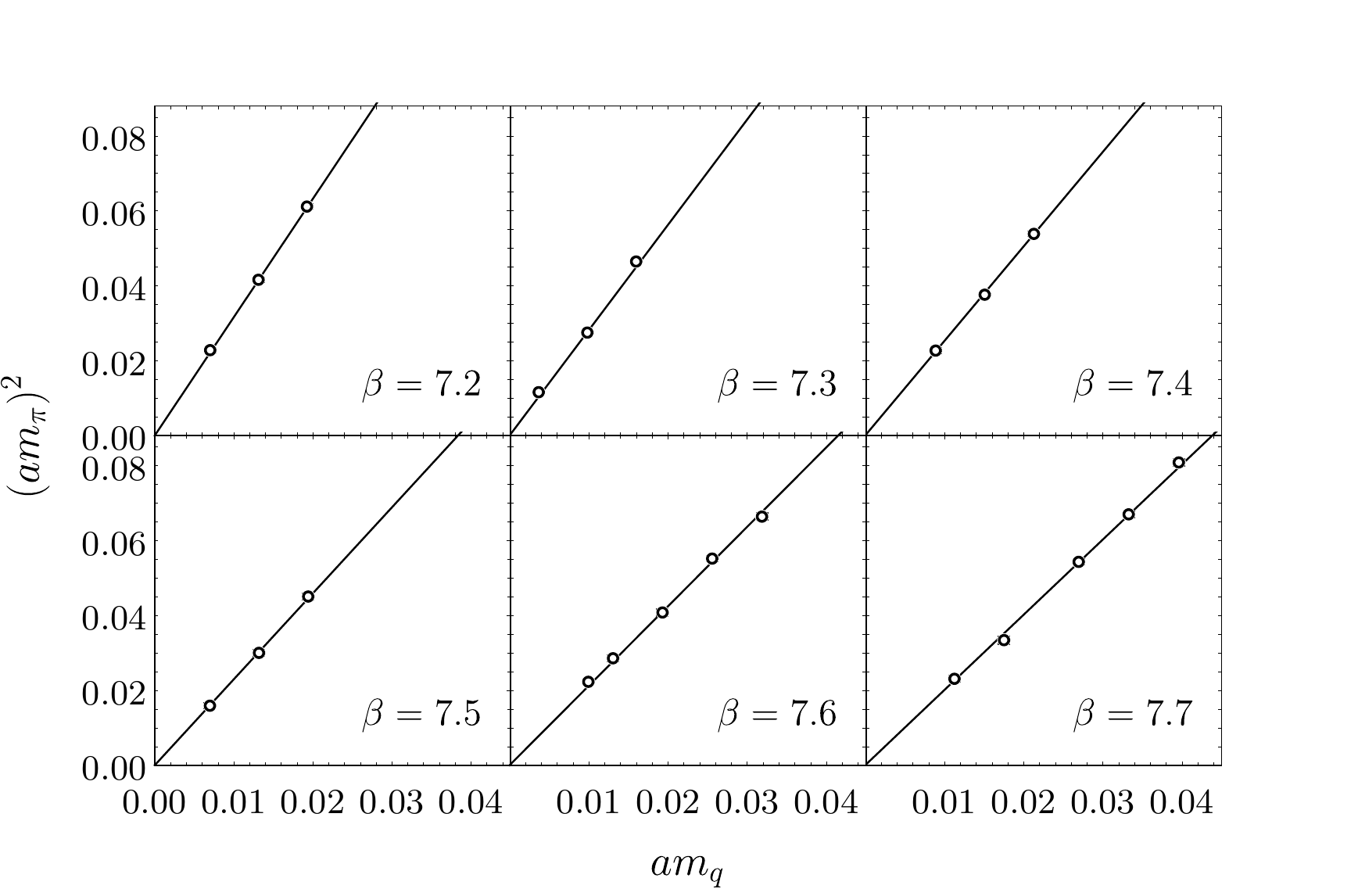}
  \caption{Pion masses measurements compared to the interpolation in Eq.~\protect\ref{eq:mpifit}. We plot the pion mass as a function of the bare quark mass defined in Eq.~\protect\ref{eq:mq}. 
  }
  \label{fig:pionmass_interp}
\end{figure}

Here we describe the parametrization for the pion mass as a function of the bare parameters $\beta$ and $\kappa$. We performed a series of fits for each set of $\beta$ values and verified that in the range of quark masses used in our simulations the leading order $\chi$PT relation $m_\pi^2\propto m_q$ describes the data well. We use a linear fit 
for the fit form
\beq
(a m_\pi)^2 (\kappa, \beta) =  \text{slope}(\beta) \left( \frac{1}{\kappa} - \frac{1}{\kappa_c(\beta)} \right) 
\,,
\eeq
to determine the critical value $\kappa_c$ where the pion
mass vanishes and the slope. Incidentally the 
quark mass for Wilson fermions require an additive
renormalization; the bare quark mass is then
\beq
a m_q = \frac1{2\kappa} - \frac1{2\kappa_c} \,.
\label{eq:mq}
\eeq
The values of $\kappa_c$ and the slope in the linear fit vary with $\beta$. We find that a qudratic fit for the slope 
and a cubic fit for $\kappa_c$ describe the data well. 
To determine the parameters, we fit this functional form to all
the datapoints in our set. We find the following interpolation
function:
\beqs 
\text{slope}(\beta)=&1.3929-1.5401\Delta\beta+1.4418(\Delta\beta)^2\\
\kappa_c(\beta) =& 0.12783-0.00370\Delta\beta\\
& +0.00375(\Delta\beta)^2+0.00135(\Delta\beta)^3 \,.
\label{eq:mpifit}
\eeqs
The interpolation is compared with the data in Fig.~\ref{fig:pionmass_interp}. We see that the interpolation
works well: most of the data points fall within 1\% from the
curve with the maximal difference at about 3\% level.

As a check, we used the scale-setting functions to find $\beta$, $u_0$, and $\kappa$ values to generate two new ensembles: one on a $28^3\times 56$ lattice with $a=0.106 \fm$ and $am_\pi=0.165$,
and the other one on a $32^3\times 64$ lattice with $a=0.093 \fm$ and $am_\pi=0.143$
.We measured the lattice spacing on them to be $0.1057(8) \fm$ and $0.0927(7)$ fm, respectively. The deviation of the central value of our measured lattice spacings from the targeted values are at the sub-percent level.
The pion mass $am_\pi$ on these ensembles was measured to be $0.1594(3)$ and $0.1379(4)$, with the deviation approximately 4\% 
between the central value of the measured masses and the target values.

\section{\label{sec:sum}Conclusion}
The nHYP smeared, clover improved fermion action with tadpole-improved L\"{u}scher-Weisz gauge fields offers an appealing option for dynamical simulations of hadron physics in QCD.
It has good scaling properties, and is relatively inexpensive to simulate when compared to other popular lattice actions such as domain-wall or overlap fermions.

We carried out simulations for 22 different combinations of $\beta$ and $\kappa$ on a fixed $24^3\times 48$ lattice. For our simulations 
we ensured the finite-volume corrections are under control (at the sub-percent level on the pion masses considered). We measure the lattice spacing using the Wilson flow parameter $w_0$ and $t_0$ and the pion mass on these ensembles. To fix the lattice spacing we use a mass-independent method by matching the value of $w_0$ in the chiral limit. We obtain smooth interpolating functions for lattice spacing $a(\beta)$, the tadpole improvement factor $u_0(\beta)$, and pion mass dependence on the bare couplings $m_\pi(\beta,\kappa)$. Interpolations based on these functions should not deviate by more than a few percent in the parameter region where the lattice spacing is between~$0.066\fm$ and~$0.115\fm$ and pion mass between~$207\MeV$ and~$834\MeV$. We expect that these
interpolations would also work reasonably well outside this range. 

As a byproduct of the scale setting method used in this paper, we also map out the behaviour of $w_0$ as a function of the pion mass.
This curve is universal, that is it does not depend on the discretization used to simulate $N_f=2$ QCD. Our determination agrees well with other calculations~\cite{Bruno:2013gha} and extends it to heavier quark masses.


\begin{acknowledgments}
  This work was supported in part by DOE Grant~No.~DE-FG02-95ER40907. RB is supported in part by the U.S. Department of Energy and ASCR, via a Jefferson Lab subcontract No. JSA-20-C0031. AA gratefully
  acknowledges the hospitality of the Physics Department at the University of Maryland
  where part of this work was carried out. The computations were performed on the
  GWU Colonial One computer cluster and the GWU IMPACT collaboration clusters.
\end{acknowledgments}


\bibliography{main}
\bibliographystyle{JHEP}

\end{document}